\documentclass[format=acmsmall, natbib=true, screen=true, nonacm=true]{acmart}

\usepackage[utf8]{inputenc}
\usepackage[ruled]{algorithm2e} % For algorithms
\usepackage{subfig}

\SetAlFnt{\small}
\SetAlCapFnt{\small}
\SetAlCapNameFnt{\small}
\SetAlCapHSkip{0pt}
\IncMargin{-\parindent}

\acmJournal{TISSEC}
\acmVolume{X}
\acmNumber{Y}
\acmYear{2022}
\acmMonth{2}
\copyrightyear{2022}
\setcopyright{rightsretained}
\acmDOI{0000001.0000001}

\usepackage[american]{babel}
\usepackage[style=american]{csquotes}
\usepackage[inline]{enumitem}
\usepackage{booktabs}

\date{February 2022}

\begin{document}

\title[Adaptive Security and Trust Management for Autonomous Messaging Systems]{Adaptive Security and Trust Management for Autonomous Messaging Systems}
\author{Habtamu Abie}
\author{Trenton Schulz}
\affiliation{%
  \institution{Norwegian Computing Center}
  \city{Oslo}
  \country{Norway}
}
\email{[Habtamu.Abie|trenton]@nr.no}

\author{Reijo Savola}
\affiliation{%
  \institution{University of Jyväskylä}
  \city{Jyväskylä}
  \country{Finland}
}
\email{savolrmx@jyu.fi}

\begin{abstract}
    With society’s increased dependence on information communication systems, the need for dependable, trustable, robust, and secure adaptive systems becomes ever more acute. Modern autonomic message-oriented middleware platforms have stringent requirements for self-healing, adapting, evolving, fault-tolerance, security, and active vulnerability assessment, especially when the internal working model of a system and the environmental influences on the system are uncertain during run-time. In this paper, we present an adaptive and evolving security approach, and adaptive trust management approach to autonomous messaging middleware systems. This approach learns, anticipates, evolves, and adapts to a changing environment at run-time in the face of changing threats. The approach combines adaptive risk-based security, trust-based security, and security-based trust: the resultant supra-additive synergy improves and increases the strength of security and the degree of trust in the system. The approach also integrates different metrics, assessment tools, and observation tools that improve and increase the assessability and verifiability of the trustworthiness of the system. Validation of results is through industrial case studies and end-user assessment.
\end{abstract}


\maketitle

\section{Introduction}

The environment surrounding modern communication and information systems is in a continuous state of change throughout the lifetime of an application. Autonomous adaptive systems deal with the uncertainty that is ascribable to a number of different factors, by being self-organizing and emergent. 

Self-organization is achieved when the system constructs, and adaptively maintains, its own behavior without external control. Emergence is the state of a system when it exhibits coherent system-wide or macroscopic behavior generated dynamically by the local interactions between the individual entities at the microscopic level. Here macroscopic refers to the dynamics of the system as a whole, while microscopic refers to the dynamics and actions of the individual entities within the system.

Message-Oriented Middleware (MOM) enables applications to exchange messages with other applications without having to know details of the other applications’ platforms and networking, thus increasing the interoperability, portability, and flexibility of architectures. Because they must be self-organizing and emergent, modern autonomous MOM platforms have stringent requirements for self-healing, adapting, evolving, fault-tolerance, security, and active vulnerability assessment; this is especially true when the internal working model of a system and the environmental influences on the system are uncertain during run-time.

GEMOM (Genetic Messaging-Oriented Secure Middleware)~\citep{Abie2008-4693668,GEMOM2007} addresses these issues and provides solutions to overcome limitations in robustness, resilience, it’s ability to evolve, adaptability, scalability, and assurance against vulnerabilities to security threats and erroneous input during run-time in the face of changing threats. In this study, we have developed adaptive trust management (ATM) and adaptive and evolving security (AES), and adaptive trust management (ATM) models that are essential to an autonomous MOM system: models that learn, anticipate, evolve and adapt to the changing environment at run-time in the face of changing threats without sacrificing too much of the efficiency, flexibility, reliability and security of the system.
The main contribution of this study is the analysis of the development and implementation of the AES and ATM models, introduced in our earlier work~\citep{Abie2009-5336915}. This (a) combines adaptive risk-based security, trust-based security, and security-based trust, whose combination improves and increases the strength of security and degree of trust in the system; this work also (b) integrates different metrics, assessments, and observation tools that improve and increase the assessability and verifiability of the trustworthiness of the system. In addition, we analyze the theoretical foundations of adaptivity, the concept all our models are based on, with its benefits and shortcomings, and introduce a trustworthiness and confidence calculation framework.

The rest of this paper is structured as follows. Section 2 gives a brief presentation of the GEMOM-enhanced secure, resilient and reliable MOM. Section 3 presents an analysis of the foundations of adaptivity with its advantages and disadvantages. Section 4 describes the AES model and the function of its core components. Section 5 describes the ATM model, which combines risk-based security, trust-based security, and security-based trust.  Section 6 shows how the AES, the ATM, and the different tool-set are combined and deals with prototyping and the validation of results. Section 7 provides a discussion on the results and novel issues raised during the development. Section 8 presents related work and a comparison of our work with that of whose work is most closely related to ours. Finally, our conclusion and future perspectives are presented in Section 9.

\section{Secure, Resilient, and Reliable MOM}

Self-healing systems attempt to “heal” themselves in the sense of recovering from faults and regaining normative performance levels by employing models, whether external or internal, to monitor system behavior and by using inputs to adapt themselves to the run-time environment~\citep{Ghosh10.1016/j.dss.2006.06.011}. Self-adaptive systems aim at anticipating changes that occur in a complex environment and automatically dealing with them at run-time based on the knowledge of what is happening in the system as guided by objectives and needs of stakeholders, and are characterized by three core functionalities: monitoring (sensing) the environment to recognize problems, making decisions on which behavior to exhibit, and realizing the behavior change by adaptation~\citep{Morandini10.1145/1370018.1370021,Laddaga4032441,Chess5386832}.

In GEMOM, the notion of fault is seen at an abstract level, and fault tolerance is looked at in a more dynamic way. GEMOM’s definition of intelligence and resilience draws attention to the fact that there can be insensitivity to faults or a low awareness of them. These faults can result in the deterioration of the functional profile of the information system, of the volumetric profile, or of the security profile. It also brings up the question of the availability of support for a reconfiguration back to an efficiently working system. GEMOM is able to rectify such vulnerability to faults by researching, developing, and deploying a prototype of a messaging platform that is evolutionary, self-organizing, self-healing, self-adaptive, scalable, and secure. GEMOM is resilient and utilizes redundant modules (hot-swap or switch-over) instantly without information loss. These resilience features allow specialist, independent system actors (e.g.,. watchdogs, security monitors, situation monitors, routers, and other optimizers) to remove or replace compromised nodes in the broader network instantly without compromising higher levels functionality and security.

 Existing MOM technologies are crude, do not scale, and are not suited to future needs. They have neither the robustness nor the resilience appropriate for future real-time systems. GEMOM provides solutions to overcome these limitations to secure autonomic messaging~\citep{Abie2008-4693668}. GEMOM is making advances in the following areas: resilience, self-healing, self-adaptive, scalability, integrated vulnerability management, better interoperability and integration of distributed systems, and holistic and systematic adaptive security monitoring and measurement.

\subsection{Resilience, Self-healing and Scalability}

The current solutions to self-healing in autonomic communications middleware are incomplete and unsatisfactory and do not support more demanding applications well. In GEMOM, resilience and self-healing are achieved by the use of an overlay of brokers that supports resilience in systems that depend on publish/subscribe MOMs, despite the lack of any privileged knowledge of the underlying infrastructure. The brokers in the overlay are called GBrokers (G-Nodes). A Broker Overlay Manager Agent (OMA) has been developed that performs autonomous adjustments to the run-time configuration of the system to preserve and maintain optimal and uninterrupted operation, also in case of partial breakdowns. It supports mechanisms for adding G-Nodes, measuring QoS between overlay components and publishers and subscribers, deciding what action to be taken to mitigate loss of QoS or breakdowns, discovering and communicating with other components in the overlay network, evaluating the performance of the system in the context of the monitored performance, establishing the state of the overlay network, and making decisions on the reconfiguration of routing and message passing.

The OMA also learns from experience and uses its new knowledge in its prediction and decision-making. Two approaches are used to achieve resilience and evolution: one being the management of reserve resources in such an overlay network, and the other being empirical correlations.

In GEMOM, scalability and resilience are achieved via cooperating brokers, publishers, and subscribers with sufficient replication of paths and namespaces, and clustering topics into groups of one or more with group replication. This allows the system to avoid overloading brokers and survive random or sudden fallout without interruption of service.

\subsection{Vulnerability Management}

The innovation in the GEMOM vulnerability management system is the integration of the detection systems, intelligent techniques, and the threat and vulnerability management tool-set into the management system. The detection systems include mechanisms for the detection of security vulnerabilities, input errors, misconfiguration errors, and bugs. The vulnerability management system includes intelligent techniques for searching and discovering vulnerabilities and other errors, detecting violations of QoS, detecting violations of the privacy policy. The threat and vulnerability management tool-set provides mechanisms for threat discovery and techniques to support generic, intelligent, adaptive approaches to robustness and security testing in a distributed environment. Knowledge of the different kinds of vulnerabilities, the software functionality, aspects of the semantics of the application domain, and protocols used is integrated into the tools to find vulnerabilities and errors in the software~\citep{Abie2008-4693668}.

\subsection{Enhanced Interoperability and Integration}

The Publish/Subscribe variant of MOM is an efficient mechanism to integrate distributed systems. This messaging paradigm provides key properties for efficient system modeling, e.g., modeling, and re-factoring of the system during run time. The message paradigm also makes the system inherently extensible as new protocols can be made by creating new topics to publish on and subscribe to. This paradigm is also a powerful base for implementation of scalability and resilience. In addition, GEMOM supports better interoperability and integration of information systems by allowing actual instances to be configured so that various functions are subcontracted to one or more separate, external or federated entities. This separation allows for a different security layout of different individual or clusters of services. For the advantages of this approach, see \citet{Abie2008-4693668}.

\subsection{Holistic and Systematic Adaptive Security}

Existing MOM systems are not able to guarantee holistic and systematic security, privacy and trust management~\citep{Abie2008-4693668}. As GEMOM advances in the areas described above, an adaptive, holistic, and systematic security approach is necessary to meet GEMOM’s stringent requirements for self-healing, adapting, evolving, fault-tolerance, security, and active vulnerability assessment. The GEMOM security solution consists of a continuous cycle of monitoring, measurement, assessment, adaptation, and evolution to meet the challenges in the changing environments. The main components are described in Sections 4 and 5.

\subsection{Autonomous and Genetic Makeup}

Some of the G-Nodes are operational nodes and some are managerial nodes. The operational nodes can be classified as producer/publisher, consumer/subscriber, and broker nodes with their specialized agents such as sensors, effectors, monitors, detectors, or analyzers. The sensors and effectors communicate with managerial nodes. Managerial nodes can be classified as QoS Mangers, Resilient Managers, Security Anomaly Managers (the Anomaly Detectors would be in the operational nodes, but there can be many and they can be layered), Adaptive Security Managers, etc. The managerial nodes make decisions about the run time operation of the system that require a wider perspective than the individual operational nodes have. Each node is aware of its context, dynamically adapting itself to continuously evolving situations, and maintains integrity by reacting to known changes, adapting to unknown changes, or dying.

The biological and ecosystem metaphors provide interesting parallels to the conceptualizations and descriptions of the G-Nodes. The overall GEMOM system architecture has a structure similar to that of a complex adaptive system that utilizes autonomous systems that mimic biological auto-immune systems at the microscopic level (operational level in this case) and that utilize the behaviors of an ecosystem of disparate entities at the macroscopic level (managerial level in this case). Biological and ecological systems maintain system integrity by reacting to known changes, adapting to unknown changes, or dying. The adaptations and responses can be at a macroscopic ecosystem level (e.g., system or species) or a microscopic biological level (e.g., molecular, cellular)~\citep{weise2008security}. Hence we can consider GEMOM as having a genetic makeup~\citep{Abie2008-4693668}.

\section{Background and Theoretical Foundations}

In the past, there has always been a problem with security in environments characterized by complexity, heterogeneity and non-uniform foundations for security tools and methods. It has not been possible to predetermine the security process nor to provide complete formalization~\citep{Shnitko2003-1222606}.  In this section, we discuss the theoretical foundations of adaptive security and trust to better understand the aspects of AES and ATM addressed in the later sections.

\subsection{Theory of Adaptation}

\begin{figure*}[htbp]
  \begin{centering}
  \includegraphics[width=0.8\columnwidth]{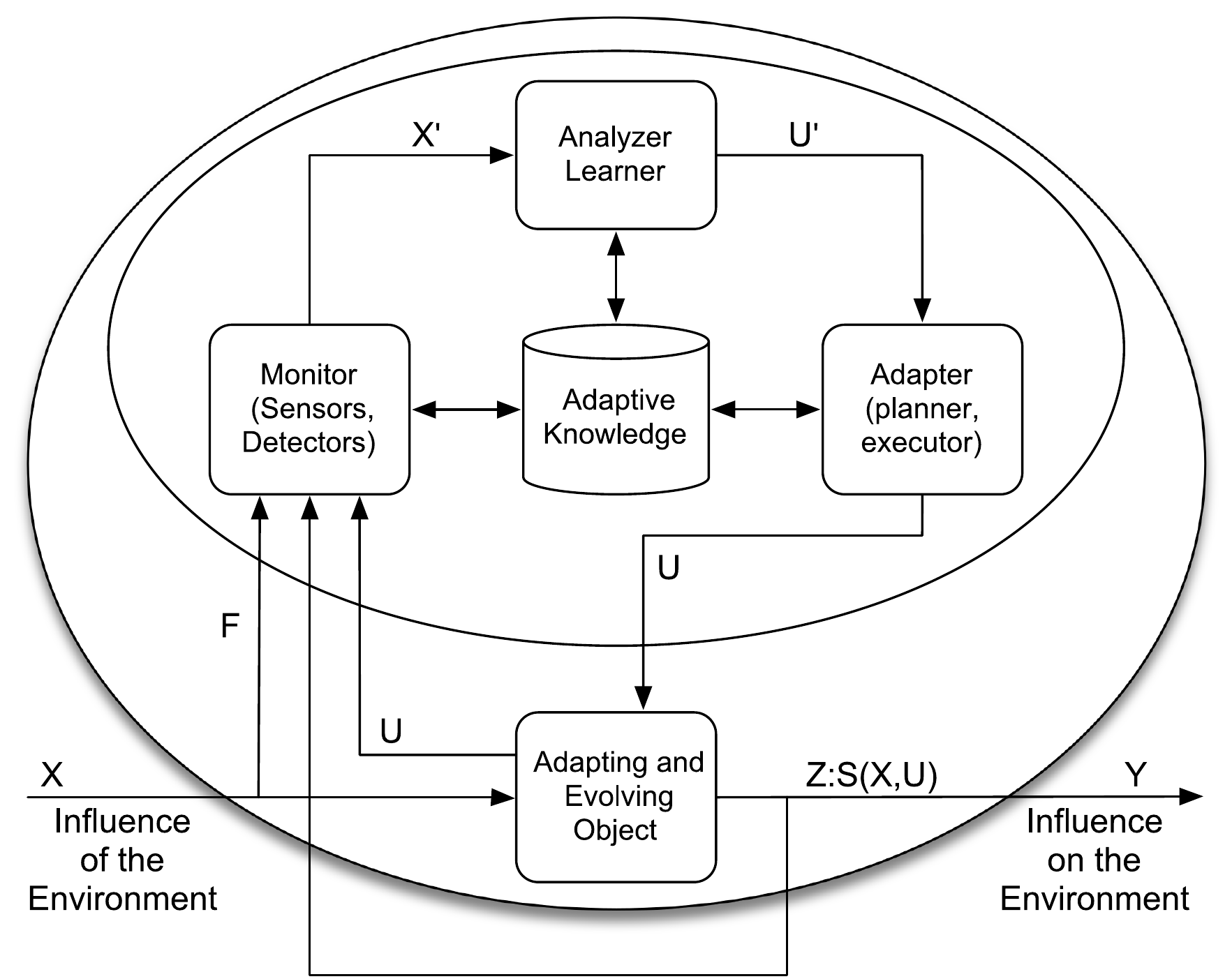}\\
  \caption{The adaptive and evolving security model.}\label{AdaptiveSecurityModel}
  \end{centering}
\end{figure*}

We used Shnitko~\citep{Shnitko2003-1222606} as the basis for the theory of our security adaptation model, a schematic representation of which is shown in \figurename~\ref{AdaptiveSecurityModel}.
Adaptation may be defined as the optimal control of
\begin{enumerate*}[label=(\textit{\roman*})]
\item specified object \textit{F} in state \textit{S} whose influence \textit{Y} on the environment is determined by the influences \textit{X} of the environment on the object, 
\item the relevant set of adaptable structures or factors \textit{U}, and
\item the goals \textit{Z} of the adaptation as defined by specified constraints on the state \textit{S} of the object.
\end{enumerate*}

Security goals are expressed as formal constraints on the state of the system, and the concepts of control theory are used to describe the dynamic security processes. The mathematical formalization and the adaptive algorithms that can learn and change their behavior by comparing the results of their actions with the goals that they are designed to achieve, are defined in Shnitko~\citep{Shnitko2003-1222606}.

\subsection{Forms of Adaptation}
Adaptation can take the form of parameter adaptation achieved by specific variations in the control parameter vector, structure adaptation achieved by dynamic changes in the structure of the system, goal adaptation achieved by formally defining specific constraints on the state of the system, or any combination of these. Adaptation can take place at any layer or across-layers, i.e., vertical cooperation among multiple system layers, horizontal cooperation among multiple platforms, and universal adaptation – combination of vertical and horizontal cooperation~\citep{samimi2004kernel}.

\subsection{Adaptive, Autonomous and Evolving Security}

\textit{Adaptive security} refers to a security solution that learns and adapts to the changing environment during run-time in the face of changing threats and anticipates threats before they are manifested. Autonomous security refers to the application of the idea of flexibility to the security space itself; this involves automating reconfiguration of the protection mechanisms, resulting in a self-protected system running without (almost) any user intervention. Evolving security refers to the modification of existing security functions and the generation of new functions for long-term adaptivity in a non-disruptive way.

Adaptive, autonomous, and evolving security involves gathering contextual information both within the system and the environment, analyzing the collected information, and making decisions through learning, responding to changes using the forms/methods of adaptation outlined above, and modifying existing functions/structures or generating new ones.

\subsection{Adaptation Technologies}

Adaptation technologies are required for static and dynamic specification of adaptation behaviors, the enforcement of the adaptation behaviors in the execution of both legacy and new applications, and the detection and resolution of conflicts. Conflicts can arise between adaptation behaviors and the execution of applications. Adaptive systems must detect and resolve the conflicts that arise due to the incompatibilities between configuration units (i.e.\ feature interactions) or due to the conflicting nature of their objectives. Current state-of-the-art resolution of runtime conflicts include methods developed using microeconomic techniques (analysis of production and pricing), and legal reasoning (logic, analysis, argumentation, and hermeneutics)~\citep{Capra2002-10.1145/605466.605472, NagwaLBadr2002ACR}.

\subsection{Special Requirements for Adaptation}

The special requirements for adaptation are that the adaptive algorithm must respond to changes in the system on the fly and the activities of the adaptive algorithm must have only minimal deviations from the normal mode of operation of the system. Additionally, adaptation must address the reconfiguration of functional logic, the architecture as a whole, and the handling of conflicts. Some obstacles to the implementation of adaptive algorithms are the complexity of the correct definition of goals and restrictions, the necessity for the on-going identification of both system and environment, and the required minimum reaction time of adaptive algorithms. Among those methods proposed in \citet{Shnitko2003-1222606} are redundancy and optimization, the usage of expert and analytical data, and special algorithms from Control Theory.

\subsection{Driving Factors and Needs for Dynamic Adaptation}

The driving factors for adaptation are:
\begin{enumerate*}[label=(\textit{\roman*})]
\item convergence of advanced electronic technologies (wireless, handheld, sensors, etc.) and the Internet,
\item the promise of instant access to data and computing,
\item the changing nature and behavior of the environment, and
\item the need for systems to operate in the face of failures and attacks.
\end{enumerate*}  

The need for dynamic adaptation is due to
\begin{enumerate*}[label=(\textit{\roman*})]
\item the heterogeneity of hardware, network, software, etc.,
\item the dynamics of the environmental conditions, especially at the wireless edge of the Internet,
\item the limited resources (such as battery life), and
\item the software adaptation technologies for detecting and responding to environmental changes, and strengthening self-auditing capabilities of “always-on” systems~\citep{samimi2004kernel}.
\end{enumerate*}  

\subsection{Fault and Intrusion Tolerance}

Fault tolerance ensures system availability by guaranteeing continuity of a service and an acceptable level of service when faults occur. Since the concept naturally lends itself to adaptability, fault and intrusion tolerance mechanisms can be used to increase the availability of a system, and previous faults perpetrated by the user can be used to increase level of suspicion. The system threat-level and user suspicion-level can be maintained by or obtained from the adaptive collectors (for example, intrusion and anomaly detectors). The user suspicion-level can also be calculated based on previous authorization and authentication events that have caused system failures or software errors. 

\subsection{Adaptive Reasoning and Decision Making Techniques}

We have investigated using the algorithm for the analysis of the dependency and correlation between features in \citet{Qu2005-1490527} as metrics to identify the minimal sets of features that must be monitored and analyzed to detect abnormal behaviors and minimize their impacts on the system operations and are basing our research on them. The model uses the following items: a feature correlation metric, a feature subset evaluation measure, a decision independent correlation defined as the ratio between mutual information and the uncertainty of the feature, a decision dependent correlation where a decision is associated with the features, a correlation measure to quantify the information redundancy between features, and a new subset evaluation measure \citep{Qu2005-1490527}, and a learning algorithm based on genetic algorithms to train the classification functions.

Our system of reasoning combines expert systems, statistical evaluations, and adaptive models (neural and fuzzy) at all levels. We apply context-aware adaptation and semi-automated reasoning to the adaptation of the security services. In areas where standard threat analysis provides specification of threatening activity patterns, the system can be directed to look explicitly for and detect those patterns. Yet, the real strength of the system lies in its own ability to learn those patterns over time, detect activity patterns that it has not seen before in real-time, and to bring these to the attention of the security analyst.

There are many optimization methods that have been proposed in the literature \citep{Peng1989-31034,Miller1993-260665,Chen2007ASO,singh2008methodologies,garcia2009anomaly}. \Citet{Peng1989-31034} presented a global optimization criteria decomposed into local optimization criteria that are used to govern node activation updating in the connectionist model for solving the problems of the simultaneous occurrence of multiple disorders that are computationally difficult. The authors also proposed what they call a resettling process to improve accuracy. \Citet{Miller1993-260665} applied genetic algorithm to the NP-hard problem of multiple fault diagnosis and compared a pure genetic algorithm with several variants that include local improvement operators that are used to accelerate the genetic algorithm in converging on optimal solutions. The authors concluded that, by using the appropriate local improvement operator, the genetic algorithm is able to find an optimal solution at orders of magnitude faster than exact algorithms. ~\Citet{singh2008methodologies} presented methodologies for optimization of distributed algorithms and middleware that consisted of:
\begin{enumerate*}[label=(\textit{\roman*})]
  \item techniques to design distributed algorithms amenable to customization, 
  \item infrastructure to analyze applications and the target platform specifications to deter-mine when customization can take place, and
  \item tools to perform the necessary code transformation and compositions to carry out the identified optimizations in the algorithms.
\end{enumerate*}
\Citet{Chen2007ASO} reviewed and summarized existing linkage learning techniques for genetic and evolutionary algorithms from three different aspects:
\begin{enumerate*}[label=(\textit{\roman*})]
  \item the means to distinguish between good linkage and bad linkage,
  \item the methods to express or represent linkage, and
  \item the ways to store linkage information.
\end{enumerate*}
  Learning linkage is the relationship between decision variables. \Citet{kramer2010evolutionary} presented a survey of self-adaptive parameter control in evolutionary computation.

\Citet{garcia2009anomaly} discussed the foundations of the main anomaly-based network intrusion detection systems technologies, together with their general operational architecture, and provided a classification for them according to the type of processing related to the “behavioral” model for the target system. The authors argued that anomaly-based network intrusion detection techniques are a valuable technology to protect target systems and networks against malicious activities.

\subsection{Advantages and Disadvantages of Adaptivity}

Adaptivity has a number of advantages in a security context. It contributes to real-world security with fuzzy definitions and under uncertain conditions. It affords access to methods and tools from Control Theory. It provides solutions to the problem of limitations in the robustness and resilience of a system and its performance. Finally, adaptivity provides a solution that learns and adapts to changing environments during run-time in the face of changing threats without significantly sacrificing the efficiency, flexibility, reliability, or security of the system. This improves the reliability, robustness, and dependability of critical systems and infrastructures.
Adaptivity has a number of potential positive impacts~\citep{samimi2004kernel,Abie2009-5336915} It increases the robustness of group communication between users with disparate devices and networks. It provides secure self-healing systems that support mission-critical communication and computation under highly dynamic environmental conditions, and self-auditing systems that report state inconsistencies, and the incorrect or improper use of components. Adaptivity allows the allocation of resources securely and dynamically in devices limited by battery-lifetime, bandwidth, or computing power. Adaptivity allows the systematic secure evolution of legacy software so that the software accommodates new technologies and adapt to new environments. Adaptivity also enables systems to operate through failures and attacks.

In the words of \citet{Hinton199-816047}: “In the trade-offs between security and performance, it seems that security is always the loser. If we allow for adaptive security, we can at least ensure that security and performance are treated somewhat equally. Using adaptive security, we can allow a system to exist in a less secure, more performant state until it comes under attack, and then we adapt the system to a more secure, less performant implementation.”

Adaptivity has some disadvantages: its effectiveness depends on the correct definition of security goals; it requires additional resources to carry out the adaptation processes, and it is not always able to ensure only minimal deviations in the system’s normal mode of operations while it is adapting.

\section{Adaptive and Evolving Security}

GEMOM has developed an AES approach to meet the requirements mentioned
above and maintain the proper balance between security and performance
in rapidly changing environments. Such an approach involves gathering
contextual information, both from within the system and from the
environment; measuring security level and metrics, analyzing the
collected information, and responding to changes. The response can be
by (a) adjusting internal working parameters – such as encryption
schemes, security protocols, security policies, security algorithms,
different authentication and authorization mechanisms, changing the
QoS available to applications, and automating reconfiguration of the
protection mechanisms – or (b) by making dynamic changes in the
structure of the security system~\citep{Shnitko2003-1222606,
  Elkhodary2007-4228616, Xiao2009article}. The analysis part of such
an approach requires flexible learning and decision-making processes
for parametrical, structural, and goal adaptation that help set
priorities and make the best decision when both the qualitative and
the quantitative aspects of a decision need to be considered.

The AES security services must adapt to the rapidly changing contexts of the GEMOM environment. The AES model consists of a continuous cycle of monitoring, assessment, and evolution to meet the challenges in the changing relationships within and between organizations both in autonomic MOM-based business environments and today’s rising threat situation. The AES model utilizes contextual information and decision making to select the “best” security model for a given situation. The AES includes the integration of monitoring, analysis functions, response functions and tool-set, elastic and fine-grained adaptive authorization, adaptive authentication, Federated Identity Management, and tools and processes for preemptive vulnerability testing and updating.

\begin{figure*}[htbp]
  \begin{centering}
  \includegraphics[width=0.9\columnwidth]{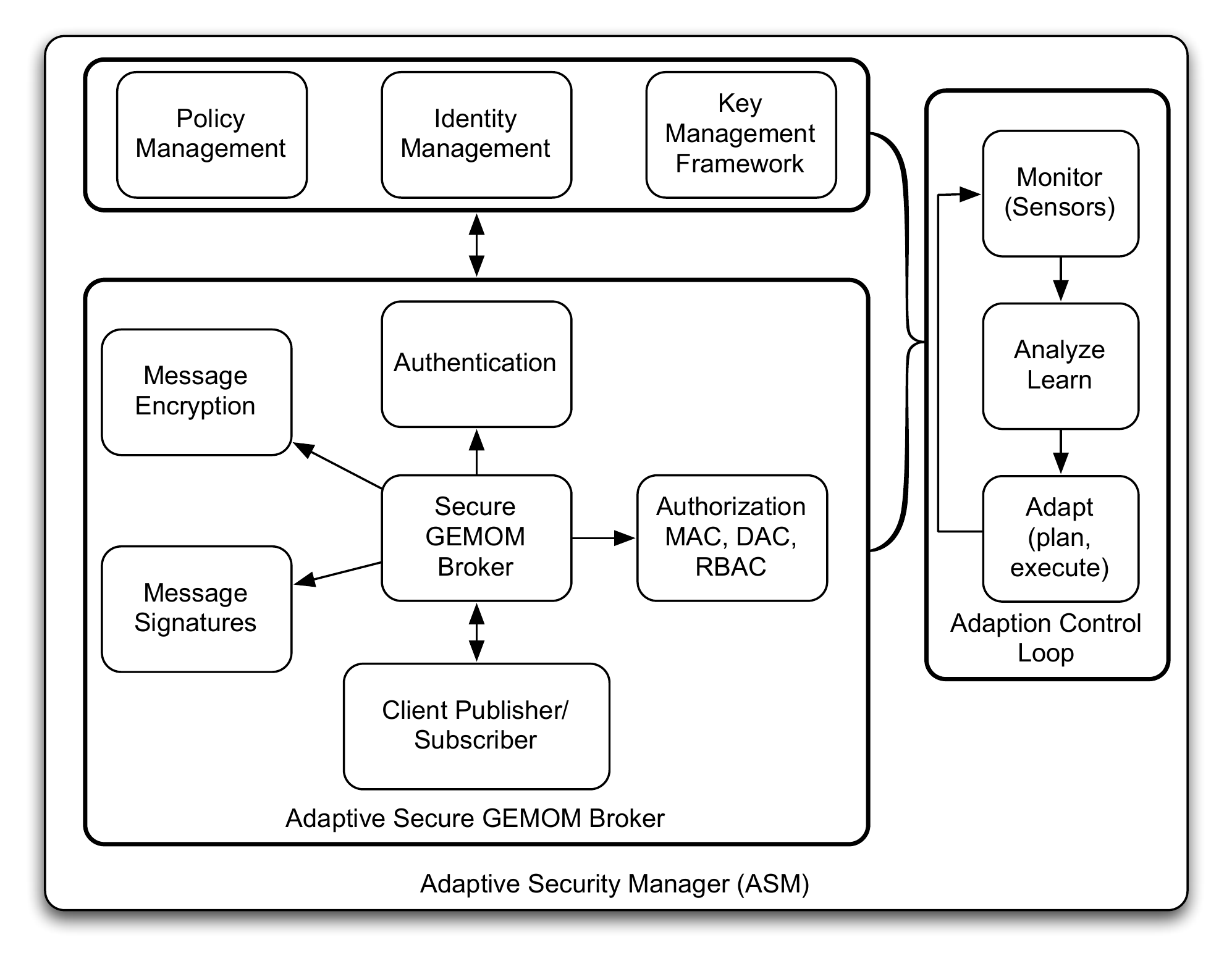}\\
  \caption{The adaptive and evolving security model~\citep{Abie2009-5336915}.}\label{AdaptiveEvolvingSecurityModel}
  \end{centering}
\end{figure*}

\subsection{Adaptive Security Manager}

The core component of the AES model is the Adaptive Security Manager (ASM), which manages and controls all the security components as an integrated GEMOM security infrastructure. \figurename~\ref{AdaptiveEvolvingSecurityModel} shows the main components of the ASM model. All these main components have been prototyped and tested.

While each component implements a local adaptation control loop, shown in \figurename~\ref{AdaptiveEvolvingSecurityModel} as the “Adaptation Control Loop”, the ASM, depicted by the outer rounded rectangle in the figure, implements a global adaptation control loop. Here the sensors are Anomaly Detector, Security Monitors, Fault Detectors, QoS Monitors, Audit, and Logging. The sensors are described below. Each component owns a public and private key pair to sign and encrypt messages, and a certificate to attest to their identities. The identity certificate, containing inter alia the component’s principal name and the name of the owner, and is signed by the Key Management Framework (KMF), which acts as a Certification Authority (CA) or Source of Authority (SOA) to guarantee the authenticity of the certificate.

The ASM implements the core functions for an adaptive secure broker, shown in Figure 2 as “Adaptive Secure GEMOM Broker”, which include Authentication, Authorization, Message Encryption, and Message Signature. For these four key components to achieve their goals, they must be supported by managerial components performing the functions of policy management, key management, and identity management. Two of the components, Authorization Module and Key Management Framework, are described below as illustrative examples in our description of the integrating architecture. 

\subsection{Authorization Manager}

Authorization in GEMOM is fine-grained and adaptive in nature. It supports access rights to clusters, groups, topics and single messages. Applying access rights to a single mes-sage is the smallest level of granularity that authorization rules can be applied. GEMOM authorization model also supports multiple user roles by defining access rights and varying performance and reliability requirements depending on the type of user.

The Rights and Permissions model in GEMOM describes the relationships between subjects, objects, roles, permissions, and constraints. Based on certain conditions, subjects have rights over objects to create, use, and delete them. Conditions specify the terms, conditions, and obligations under which rights can be exercised. 

Operations in the GEMOM system could be performed based on certain extended composite keys. These extended composite keys consist of any combination of the following sub-keys:  user, strength of authentication, context, time when operation is per-formed, and security profile of the system. The context can be an environmental context (e.g., bandwidth, stability of connection, power of the local machine), access context (e.g., include descriptive justification of the access operation, where and when the requested data goes, the duration of the use of the data, the precondition and post-condition of the access operation), or the business context (e.g., in investment banking the same person using the system for trading or risk management implies a marginally different profile).

It is the pair (Actor, Authentication Strength) by which the actor is authenticated that is a unit-entity that GEMOM authorizes. With this pair as a basic composite key the GEMOM authorization process proceeds by using the following key properties: (a) a user belongs to a group, and the basic user authentication strength key is translated into a vector of group authentication strength pairs; (b) the system is perceived as having certain multi-dimensional security profile, and boundaries are defined in each dimension; (c) an application is divided into an arbitrary set of modules, and an abstract notion of operation on a module is defined where a module can allow an arbitrary number of operations to be performed on it. Access rights are defined to the pair (module, operation); and (d) certain groups of users that are authenticated with strengths that fall into certain ranges are allowed to perform certain operations on application modules within certain periods of time, within defined context boundaries, and within certain dynamic security boundaries. The development of adaptive features of the authentication, identity management and authorization processes are described in detail in \citet{Abie2008-4693668} and \citet{Abie2009-5336915}.

The development of trust in GEMOM involves the identification and understanding of the risks and vulnerabilities of the GEMOM system and forming trust solutions to address the risks and vulnerabilities. Trust building by allowing gradual establishment of trust based on attributes or credentials and by using the authentication strength described above is integrated.

As a flexible and adaptive authorization processes in the GEMOM system, the model supports multiple and dynamic user roles, federated rights, the specification of fine-grained users, messages/topics, and access rights. The system is flexible enough to allow an independent specification of each user’s access rights for each topic or group of topics. The adaptive authorization performs access control by making access decisions based on security policies and permissions and rights specified by the Security-Token described below. It provides adaptive authorization through changing security policies, algorithms, protocols, and encryption schemes according to context parameters, such as environment, system threats, user threats, usage, metrics, faults, or quality of service (QoS). This is further elaborated in Section 6. 

\subsection{Key Management Framework}
The Key Management Framework (KMF) ensures secure communications and message delivery in GEMOM, and is based on \citet{Pallickara2006-4100475}. It performs several core functions:
\begin{enumerate}
\item It generates secret symmetric keys for encrypting and decrypting topic payloads on secure topics.
\item It maintains the list of authorized clients associated with a secure topic, and maintains authorization information related to each of these clients that may be registered to publish, to subscribe, or both.
\item It generates security tokens for every authorized entity. This security-token establishes an entity’s rights for a topic and the duration of the validity of these rights. To enable tampering-evidence the contents of this security-token are hashed and signed by the KMF.
\item It securely distributes secret keys associated with a secure topic by wrapping the secret key using the entity's public key. Entities can register new secure topics or request keys for already existing secure topics.
\item It sets up topics for entities to communicate with the KMF. This was original plan was to use a discovery node, but this is currently done by communicating on well-known topics.
\item It listens for changes from the security monitor and invalidates current secret keys for secure topics and regenerates keys for higher or lower encryption levels. For ex-ample, the KMF could generate stronger keys when it suspects that the system is under attack or it could generate weaker keys when networking or processor capacity is an issue.
\end{enumerate}

The KMF is divided into two parts: A server that acts as a special node and handles all the administration and encryption duties, and a client library for making KMF tasks easier for clients. A given KMF may manage more than one secure topic, but a given secure topic can be managed by only one KMF. Clients communicate with a KMF by sending messages to well-known topics that the KMF server listens to. Messages for registering a secure topic or requesting a secure topic’s key are encrypted with the KMF’s public key. The response sent by the server is encrypted using the client’s public key. For security token requests, the security token is signed by the KMF for validity.

The secret keys generated by the KMF are currently using AES and can be from 128-bit to 256-bit in size. If the KMF finds cause, it will invalidate all its currently issued keys and regenerate new ones. All clients that use secure topics listen to invalidation information from the KMF regarding invalidating keys. In the current implementation, clients must request new keys for the secure topics after a key has been revoked.

Clients can sign the messages that they publish. This is useful for clients that want to ensure that the data they are seeing is actually being published from the correct source. To sign data, the publisher uses its private key to generate a signature of the current data and then adds a new field to the data that includes the signature. Other clients can then use the publisher’s public key to verify the data. There is no official topic for publishing public keys at the moment, so clients wishing to use this capability must provide their own mechanism for exchanging public keys (e.g., communicating over a private topic).

\textit{Tokens} are essential for protecting messages (e.g., subscriptions) from selective drop-ping DoS attack. A token is a pseudonym for a topic name and the concept of per-topic key for achieving confidentiality and integrity has been investigated. Signatures play a pivotal role in achieving message authentication and protecting the publish/subscribe services from flooding-based DoS attacks.

\subsection{Self-Protection}
\label{sec:self-protection}

A self-protecting system, as defined by IBM~\citep{IBM2005}, can anticipate, detect, identify, and protect itself against threats, unauthorized access, and denial of service attacks. Therefore GEMOM as an autonomic MOM has to implement self-protecting capabilities that can detect hostile behaviors as they occur and take corrective actions to make it less vulnerable. In GEMOM, the self-protection is handled by a single entry point (micro property) that gives each node authorization, a coordinated defensive group attack by the other nodes (a macro property), or a combination of the two (defense-in-depth). 

Intrusions can be handled by triggering a one-shot behavior of the GEMOM system. Yet, the GEMOM system has to be alert, so the degree of protection over time (ongoing) is important. Based on the structure, the AES self-protection can be decomposed into three levels (threat points) whose granularity is summarized in \tablename~\ref{table:1}. The three levels work together to achieve the necessary self-protection of the GEMOM system.

\begin{table}
\caption {Self-Protection Threat Points and Solutions}
\label{table:1}
\begin{tabular}{p{3cm}p{3cm}p{6.3cm}} 
  \toprule
  \textbf{Granularity level} & \textbf{Self-protection} & \textbf{Self-protection} \\
  (\textbf{threat point}) & \textbf{Issues} & \textbf{Solutions} \\
  \midrule
 Communication or network level  & Protection from malicious node & Network level self-protection mechanisms.
 
Network level trust management scheme.

Confidentiality, integrity or authenticity of underlying IP-network can be guaranteed using TLS/SSL connection between routing nodes.

Trust models at this level help assess the quality of new joining nodes and the degree of confidence in their behaviors.

Anomaly-Based Self-Protection~\citep{qu2006anomaly}.
 \\
\midrule
Broker Nodes level & Protection of run-time environment & Trusted execution environment for nodes.

Node self-protection such as mutual authentication and authorization of broker nodes for accurate namespace resolution to protect against threats from rogue brokers and to protect confidentiality and integrity.\\
\midrule
Publisher/subscriber level & Protection from malicious publisher/subscriber & Security contracts or service level agreements. 

Use of authentication and sub-set of mechanisms to enforce access control for authorized publishers/subscribers.

Node-level trust management scheme such as certificate- or token-based, and adaptation and maintenance of the trust level over time by building a reputation feedback mechanism.\\
\bottomrule
\end{tabular}
\end{table}

\subsection{Security Monitoring}

The security and QoS management of GEMOM is based on monitoring utilizing appropriate metrics \citep{Savola2010article}. The monitoring functionalities were developed and validated against different scenarios of a changing environment. A research prototype Monitoring Tool was developed, supporting both the security and QoS management. The tool includes metrics, thresholds, and sufficient mechanisms for the collection of evidence. The QoS Manager and Security Manager do the actual measuring and make decisions, while the Monitoring Tool carries out tasks like the management of measured data, threshold, averaging, and metrics aggregation. The monitoring solution allows con-figuring different manager tools to be used together with the Monitoring Tool. Some of the sensing in the GEMOM system is done via QoS measurements. 

The QoS Manager and Resilience Manager resolve problems in their domains by using QoS measurement. Details of the Monitoring Tool implementation can be found from Savola and Heinonen~\citep{savola2010security}. Figures 3 and 4 show screenshots of the metrics management view with examples of authentication and access control metrics categories and the requirement for the minimum authentication strength at different time instants, respectively.
\begin{figure*}[htbp]
  \begin{centering}
  \includegraphics[width=0.8\columnwidth]{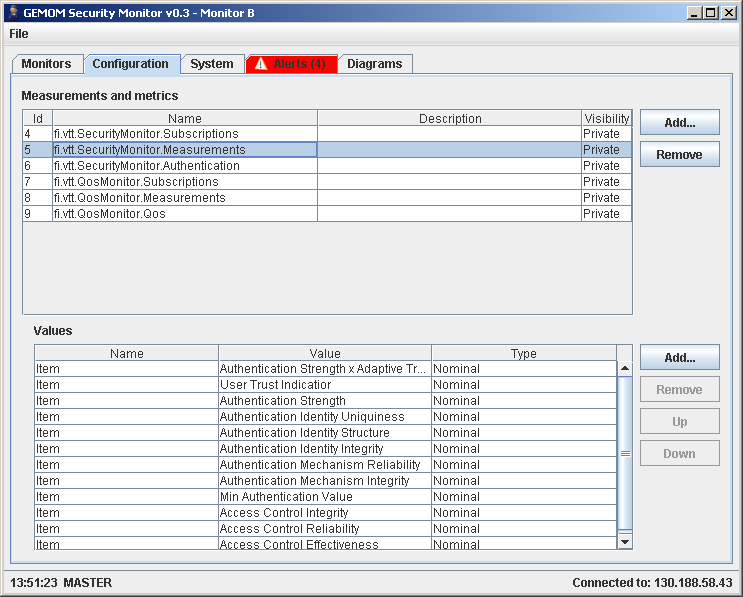}\\
  \caption{The GEMOM Monitoring Tool’s authentication and access control metrics categories.}\label{MonitoringToolAuthentication}
  \end{centering}
\end{figure*}

\begin{figure*}[htbp]
  \begin{centering}
  \includegraphics[width=0.8\columnwidth]{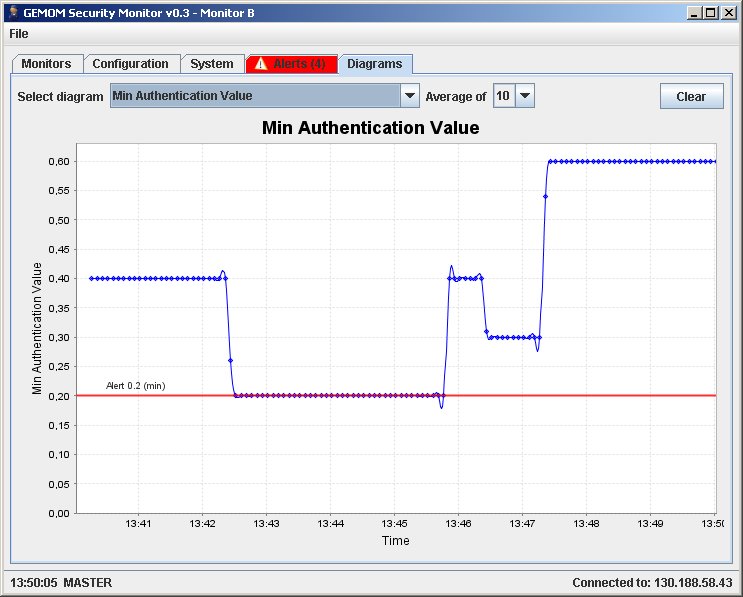}\\
  \caption{The requirement for minimum authentication strength at different times in an example monitoring case.}\label{AuthenticationRequirement}
  \end{centering}
\end{figure*}

Combining on-line and off-line metrics supports adaptive measurement: results from applying the off-line metrics are used to reconfigure the on-line metrics. The combination of on-line and off-line metrics is also used to establish an operational feedback loop to risk analysis activities. The detection of anomalous and the monitoring of behavioral patterns (reactive operations), and the use of up-to-date information about threat, vulnerability, and reputation levels are processed using off-line metrics (proactive operations) based on the online metrics. This metric combination enables us to cope with changing threat levels and develop the system further so that it can achieve and assure security over time~\citep{Savola2010article,Savola2009OnLineAO}. The discovery of anomalies, i.e., patterns that are anomalous to constructed (learned) models of normal characteristics, is the task of the Anomaly Detector. Learning is the task of the Profiler component. While QoS data is fed to the QoS Manager; the Adaptive Security Manager; and the Security Measurement Manager, the output from the Anomaly Detector is fed to the Security Measurement Manager.

\subsection{Security Metrics and Their Development}

Adapting to changing operational environment – including threats and vulnerabilities – during run-time requires metrics that provide sufficient and credible evidence for decisionmaking. These metrics should be
\begin{enumerate*}[label=(\textit{\roman*})]
  \item accurate,
  \item meaningful, and
  \item measurable.
\end{enumerate*}    

The correct and risk-aware Security Objective (SO) management is vital for security metrics development, management, and setting the reference level. The resulting security metrics can be only as effective as the questions they answer, which are based in a meaningful way on SOs. It is advantageous to utilize hierarchical metrics development and maintenance decomposition of SOs increases meaningfulness by explicitly showing the relationship between the high-level questions and the metrics and measurements. It also lowers the bias between rational measurements and heuristics-based decision-making. 
Especially in security metrics used to depict the effectiveness of security-enforcing mechanism should support prediction. Predictive correctness is one of the main challenges in these kinds of security metrics.

A general and systematic security metrics development method based on SO decomposition was introduced in Savola and Abie~\citep{Savola2010article}. The method is several highly iterative steps: starting from threat and vulnerability analysis and resulting to a detailed and balanced collection of security metrics. The method also integrates QoS metrics. The method supports well-defined metrics, since the relations between high-level requirements and low-level metrics are shown and maintained using the decomposition approach. Savola and Abie~\citep{Savola2010article} show example metrics developed based on Basic Measurable Components (BMCs), the leaf components of decompositions developed for the purposes of adaptive security management in GEMOM.

SOs should be developed in a prioritized way, based on high-quality risk analysis re-sults. The risk-driven security metrics development is a top-down activity. Yet, bottom-up thinking is needed too: metrics are worthless if the information to be measured is not available or attainable. In GEMOM, security measurability is enhanced by parallel, “hand-in-hand” development of the metrics and the measurement architecture. The mechanisms in GEMOM include utilization of the publish/subscribe mechanism for measurement purposes, auto-recovery on error, measurement mirroring, data redundancy, multi-point measurement, integrity and availability checks, a systematic timing concept of the measured data, use of shared metrics and measurement repositories, utilization of QoS, performance, delay, packet loss rate parameters, and other indicator data for the purposes of security metrics~\citep{savola2010security}.

\section{Adaptive Trust Management}

\begin{figure*}[htbp]
  \begin{centering}
  \includegraphics[width=0.8\columnwidth]{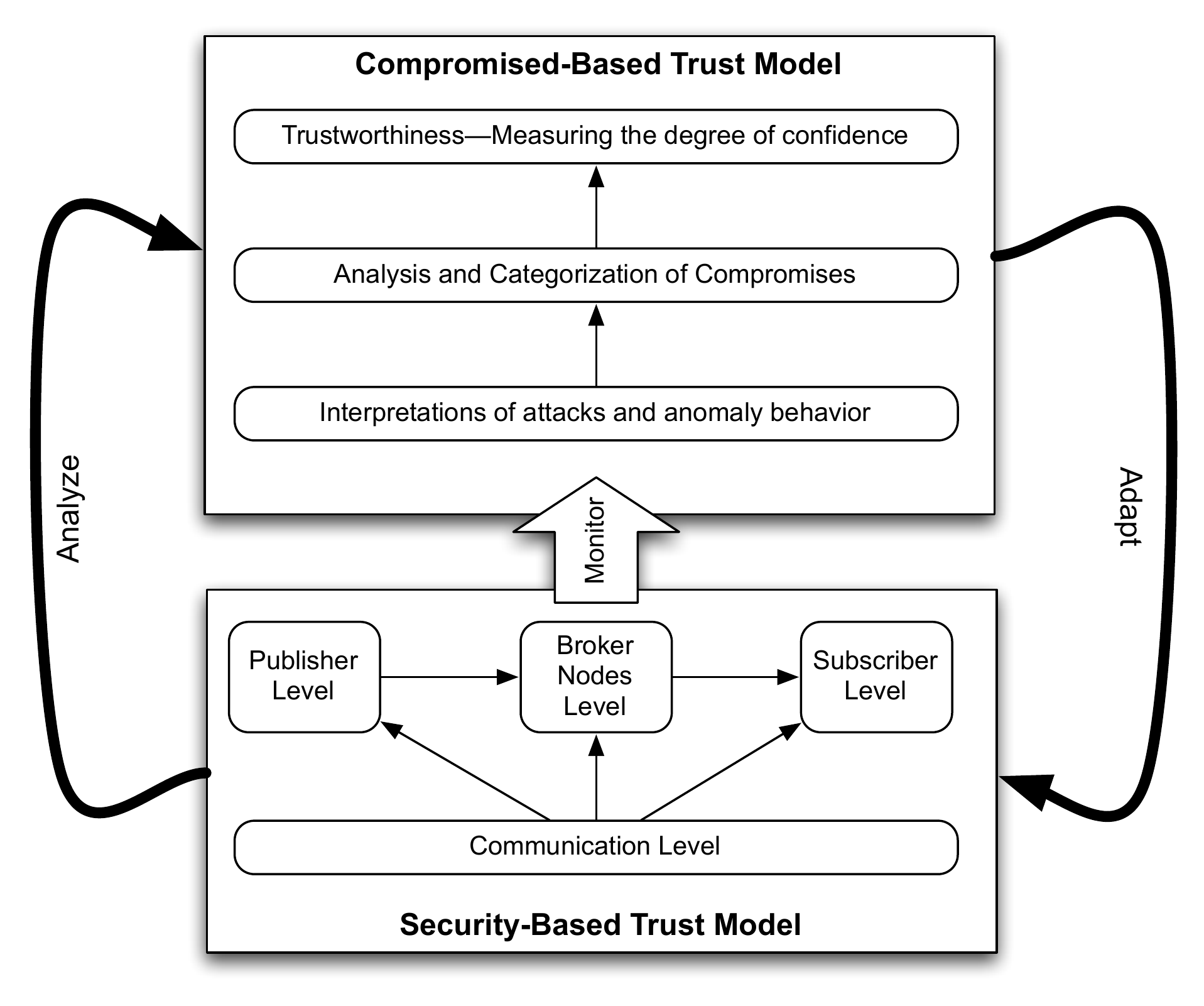}\\
  \caption{The adaptive trust management model.}\label{AdaptiveTrustManagementModel}
  \end{centering}
\end{figure*}

The GEMOM ATM model is logically organized into a security-based model and a com-promise-based model. These two models work together to achieve the adaptability of trust and security in GEMOM. \figurename~\ref{AdaptiveTrustManagementModel} shows these two models and the interaction between them. Our compromise-based trust model is inspired by the recognition of the crucial role played by the assessment and management of trust and by the rejection of the assumption that trust relationships are binary and static in nature and that models based on such an assumption are good approximations to real-life computing situations, expressed in \citet{shrobe2000active}. The Security-based trust model is achieved via the security services of the AES, which supports the establishment of trust through the provision of a secure and trustworthy environment.

By concentrating on the notion of a fault, the GEMOM project expects to make advances in the security of messaging. In addition to understanding intuitively the nature of a fault that stops any actor being operable, leads to a connection being lost, etc., the GEMOM project extends the notion of fault to include compromised security or the unavailability of adequate bandwidth in the first iteration. The final iteration also includes the abstract notion of a compromised-resource. Therefore, we developed a compromised-based trust model for GEMOM based on \citet{shrobe2000active}. Our trust model provides information about any attack on the system and the nature of that attack for the purpose of establishing whether, and if so how, different properties of the system have been com-promised. In addition, it establishes whether these properties can be trusted for a particular purpose even if it’s compromised and to what degree these judgments should be suspected or monitored. The trust model is organized into three levels and the three levels work together to achieve its adaptability.

The overall ATM system is designed to adapt to the dynamism of the GEMOM environment and to changing degrees of risk of compromise in the GEMOM components. The ATM does this by deciding dynamically which approach is to be adopted, and which approach provides the best likelihood of achieving the greatest benefit for the smallest risk, i.e., maximizing the value of taking a risk. Consequently, our adaptive risk-based security adapts its decision based on the computation of the impact of the security risk.

\subsection{Risk, Trust and Security Adaptation}

Risk is an inherent part of any security or trust system. To ensure the secure operation of a system, it is necessary to have some well-founded practice for the identification of security risks (as well as the application of appropriate controls to manage risks)~\citep{AbieRiskAnalysis,Abie2005aconceptual, trvcek2010advanced}. Risk adaptive access control is an important emerging technology that determines access based on a computation of security risk and operational need, and adapts its decision thresholds such that operational need can trump security risk when appropriate~\citep{MCGRAW2009}. Risk ranking can also be used to evaluate various alternatives by com-paring the risk associated with them (taking into account deterioration rates, relative frequency of overload, costs of failure, costs and efficiency of repair strategies, impacts of compromise, etc.), and is based on an overview of the currently most popular dangerous dangerous types of attack. The result of risk ranking is the provision of a common, updateable, and collaborative framework for updating the security vulnerabilities of a system and validating it, which serves the interests of service providers and users alike. 
Trust is a necessary prerequisite basis for a decision to interact with an entity. Trusting an entity is always associated with risk as there is always a chance the entity will behave contrary to expectations. Security is the reduction of risk to an acceptable level through the use of enforcing mechanisms where what constitutes an acceptable level is decided by the extent we trust the function of the system. Trust reduces risk, builds confidence in the value of a business, and provides security. Security supports the process of establishing and maintaining trust by provisioning a secure and trustworthy environment. Security reduces the rate and severity of compromises by continuously adjusting and responding to emerging and changing threats.

Figure 2 in \citet{Abie2010SelfhealingAS} shows the relationships of cause and effect as a foundation. The three emerging areas of adaptive risk-based security, trust-based security, and security-based trust can form a combination. This supra-additive synergistic effect improves and increases the strength of the security and the degree of trust in the system, and reduces the rate and severity of compromises by continuously adjusting and responding to constantly emerging and changing threats.

\subsection{Interpretations of Attacks and Anomalous Behaviors}

At this first level, the necessary information is collected, filtered, and organized for the purpose of triggering analysis and inference. It is not the precise nature of the attacks and anomalous behaviors that have taken place that is of primary interest, but their value as an indication of how the system may have been compromised. It is our intention, on the basis of the work done by the Intrusion Detection System (IDS) community and others, to produce annotated taxonomies of different types of attack and to determine how they can be fed into the next level of the trust model~\citep{shrobe2000active}. As part of this work, we developed a holistic framework for security metrics development based on threat and vulnerability analysis, security requirements, and use case information~\citep{Savola2010article}. The adaptive security tool-sets developed in the project are used for the interpretations of attacks and anomalous behaviors.

\subsection{Analysis and Categorization of Compromises}

At this level we address the following three issues: 
\begin{itemize}
\item Standard security protections (privacy, integrity, authentication, authorization, confidentiality, non-repudiation, etc.) and operational properties such as QoS are used to classify compromises. For this purpose, we have developed a framework for the identification of basic measurable components, and a process for the development of security metrics~\citep{Savola2010article}. 
\item Control properties – DoS or other compromises of the system – such as
  \begin{enumerate*}[label=(\textit{\roman*})]
  \item the degree of confidence,
  \item the observability of security and operational properties, and
  \item the degree of control (loss of observability)
  \end{enumerate*} – can come about through attacks on the GEMOM monitoring system. These are observed by the adaptive toolsets described above. There is a need to schedule certain systems for the recognition of low-level attacks with end-to-end and real-time requirements to guarantee their instantiation~\citep{taylor2003attack}. 
\item Distinguishing different sub-entities – G-Nodes communication network, the operational processes, the security mechanisms, more finely to distinguish the types of in-formation, operations, information sources, or destination affected, and the identity and roles of entities participating in – or affected by – the compromises.

\end{itemize}

\subsection{Calculation of Trustworthiness and Confidence}

At this level, we measure the degree of confidence in the security mechanisms used: in the interpretations of attacks and anomalies, in the analysis and categorization of possible compromises and their possible impact, in our ability to make rational decisions, in the adaptation achieved, and in the maximization of the value of taking risk. For this, we collect knowledge of (a) attack types – to guide our attempts to defend against future attacks, (b) compromises – to indicate the threats to operations, and (c) trust – to state guides as to how GEMOM carries on in the face of partially understood compromises. 

For the measuring and calculating the degree of confidence, we have proposed, discussed, and developed a flexible framework for the assessment and calculation of the degree of the trustworthiness of and confidence in the measurement of the overall security level of the system as a whole~\citep{Savola2009OnLineAO}. The framework is dynamic and adaptive, depending on the behavior and security measurement results of the measurable components. The main contributions of this framework are three sub-frameworks,
\begin{enumerate*}[label=(\textit{\roman*})]
\item for the calculation of levels of security, trust, and confidence; 
\item for mapping trust and confidence into a trustworthiness metric; and
\item for the assessment and calculation of the trustworthiness of the measurements of the overall security of the system through the combination of risk-based security, security-based trust, and trust-based security.
\end{enumerate*}
  These frameworks are based on our earlier framework~\citep{Savola2010article}, but this time we separate trust and confidence, and then combine them to form a trustworthiness metric. The definitions of trust, confidence, and trustworthiness are similar to those presented in \citet{Zouridaki10.1016/j.adhoc.2008.10.003}. The difference is that we used trust to mean the extent we trust the reliability of our estimation of the security level of each BMC (Basic Measurable Components) and used confidence to mean the measure of the level of the accuracy of or the assurance in this trust relationship. The values of trust and confidence are both expressed as a number between zero and one based on Bayesian statistics, where a trust value equal to one indicates absolute trust and a value close to zero indicates low trust. Similarly, a confidence value equal to one indicates high confidence in the accuracy of the trust value and a value close to zero indicates low confidence. To facilitate trust-based decisions, trust and confidence have been combined into a single value – trustworthiness – whose value is measured in the same way. For schematic diagram and the mathematical foundation of this calculation see \citet{Savola2010article}.

The assessment and calculation of the trustworthiness of the measurement of the security of the system as a whole is based on the aggregation and propagation of the different measurements made of the system at different levels. This propagation and aggregation can for the most part be automated since the interdependencies of the BMCs are modeled automatically on the basis of the structural and functional relations between them.

We assess the trustworthiness of and confidence in the overall GEMOM system according to this trust model to make decisions about how to adapt to rapidly changing environments. The trust model gives a probabilistic representation of the trustworthiness of and confidence in each measurable component in the system.

\section{Adaptive Integration Architecture, Prototypes, and Validation of Results}

As already stated, the AES includes adaptive integration functions and tool-sets. This section briefly describes the integration of these tool-sets using adaptive authorization as an example of how these tools can be integrated. 

\subsection{Adaptive Integration Architecture}

\begin{figure*}[htbp]
  \begin{centering}
  \includegraphics[width=0.9\columnwidth]{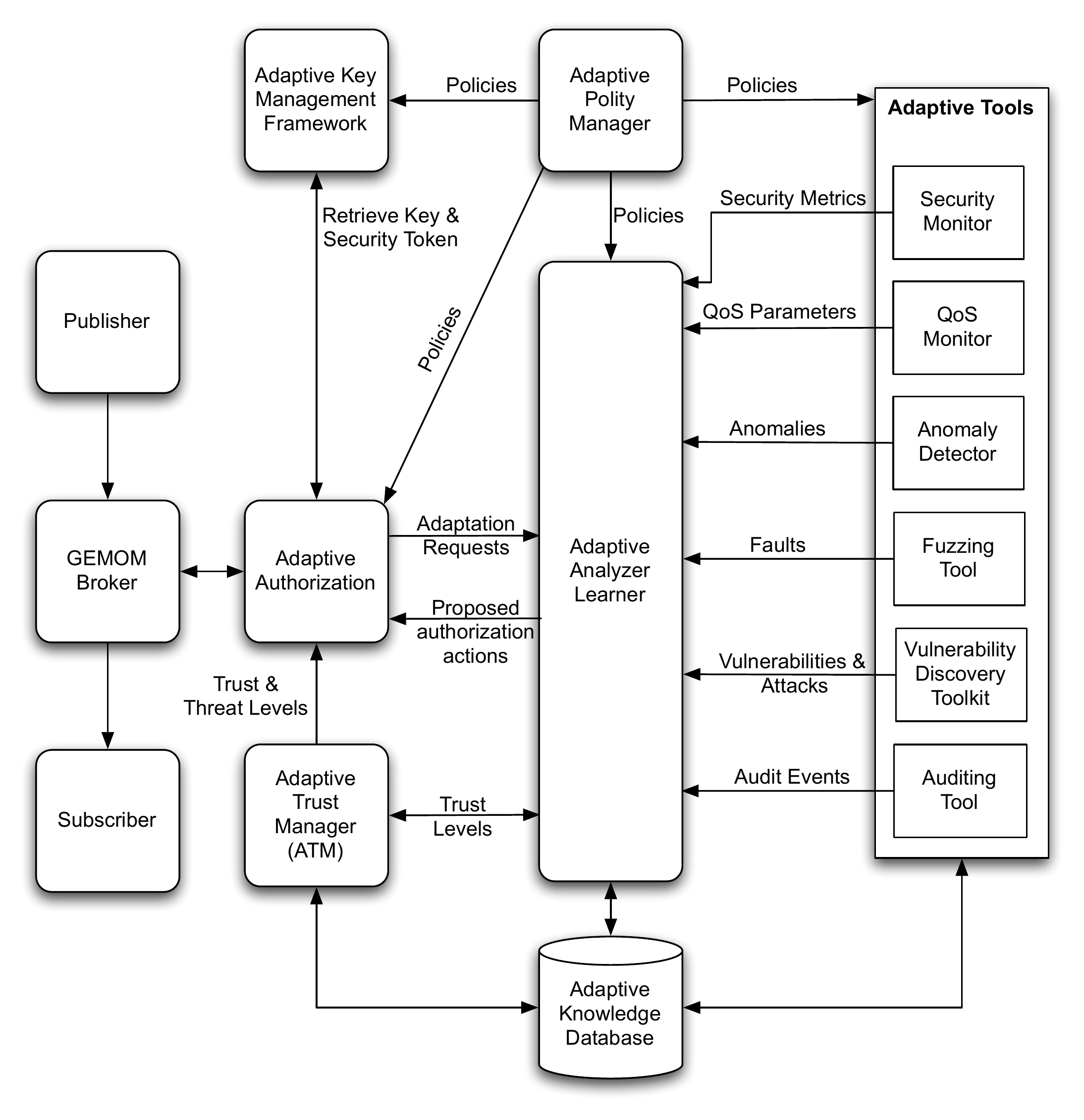}\\
  \caption{The adaptive integration architecture tools.}\label{AdaptiveIntegrationArchitecture}
  \end{centering}
\end{figure*}

\figurename~\ref{AdaptiveIntegrationArchitecture} shows the adaptive integration architecture. All these tool-sets, with the exception of the last two mentioned under “Adaptive Tools,” have been prototyped and tested. 

The Adaptive Authorization component provides adaptive authorization through changing security policies, algorithms, protocols, and encryption schemes according to context parameters such as environment, system threats, user threats, trust levels, usage, security and trust metrics, faults, and quality of service. Fault and intrusion tolerance mechanisms are used to increase the availability of a system, and previous faults caused by the user are used to increase suspicion-level. The system threat-level and the user suspicion-level are maintained by and obtained from the Adaptive Tools (Security Monitor, Anomaly Detector, Fuzzing Tool, etc.). The Adaptive Authorization component allows trust building by allowing the gradual establishment of trust based on attributes, credentials, identities, anomalies, and attack-based trust models~\citep{Abie2010SelfhealingAS}.

In its adaptive form, the Adaptive Authorization component associates authorization policies and security tokens with trust and threat levels to adapt its restricting or relaxing of authorization constraints. \figurename~\ref{AdaptiveIntegrationArchitecture} depicts the relationships between the adaptive authorization component and other security components.

The Adaptive Analyzer component analyses the collected information using established analysis and decision-making methods. It processes the collected data, along with other information (e.g., security policy, threat levels, or trust levels boundaries) and proposes actions to bring about a new stage. The Adaptive Tools sense and gather contextual information from within the system and from the environment, and they distribute information about the security environment to the Adaptive Analyzer and adaptive database. The Vulnerability Discovery Toolkit allows the identification and understanding of the risks and vulnerabilities of the GEMOM system and the forming of trust solutions to address the risks and vulnerabilities. The Fuzzing Tool allows an effective black box testing technique to be used for finding security flaws from software.

Policy dissemination: Another major requirement for our ASM implementation is that it should be able to disseminate authorization policies through the publish/subscribe system in a semi-automated way. When a particular policy, such as a subscription or advertisement policy for a topic, needs to evolve, the policy owner (e.g., the topic owner) publishes a policy evolution topic. These topics never reach publishers and subscribers nodes, but they will reach every Node that might currently be caching the policy for this topic type. This is implemented by the Adaptive Policy Manager.

\subsection{Prototypes}

We used prototyping to explore design alternatives, test theories, and confirm performance. We used our experience to tailor the prototype to our specific requirements. Our prototypes are being used both to confirm and verify user requirements that our design must satisfy through case studies, and to verify the performance and suitability of our design approach. Following the common strategy of the GEMOM project, i.e., design, test, evaluate and then modify the design based on the analysis of the prototype, we have developed, prototyped, and lab-tested: a full-featured message broker, transparent completion and encapsulation publishing framework, adaptive security implementation (authentication, authorization, key management, identity management), a MOM Intelligent Fuzzing Tool for a pre-emptive security black box testing, a Security Monitoring Tool, and tools for the management of configuration and deployment and development process.

\begin{figure*}[htbp]
  \begin{centering}
  \includegraphics[width=0.8\columnwidth]{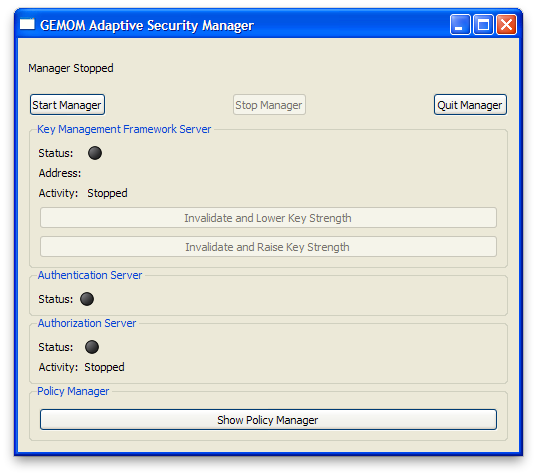}\\
  \caption{The GEMOM adaptive security manager controls the policy, authentication, authorization, and KMF.}\label{AdaptiveSecurityManager}
  \end{centering}
\end{figure*}

\begin{figure*}[htbp]
  \begin{centering}
  \includegraphics[width=\columnwidth]{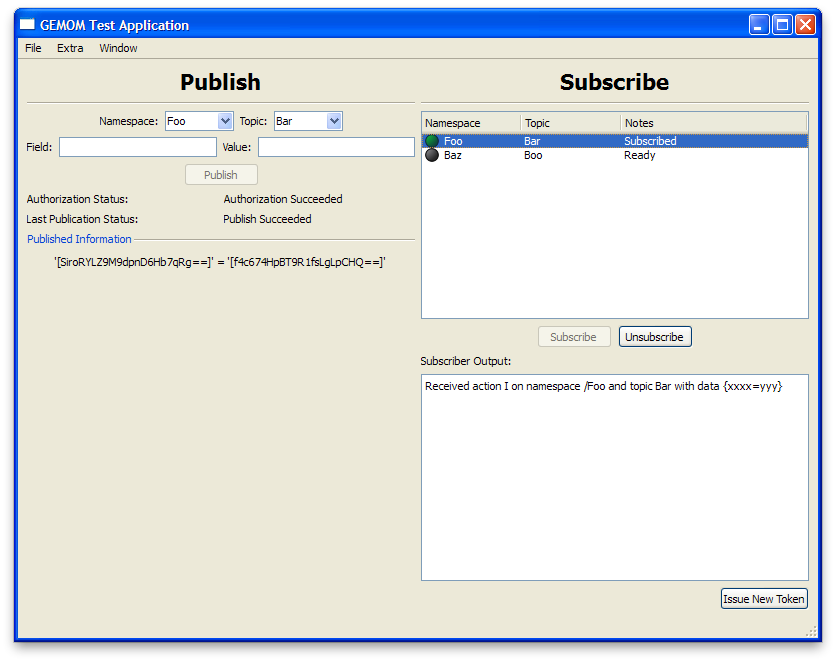}\\
  \caption{An authenticated GEMOM client publishing encrypted data using a key from the KMF; the left side shows the publicly viewable information, while the subscriber output on the right is decrypted.}\label{GEMOMTestApplication}
  \end{centering}
\end{figure*}

We have also developed demonstrators for enhanced resilience, QoS and security implementation, security and QoS monitoring system, integrators for well-known commercial MOM systems (JMS, Tibco’s, Reuters, and IBM’s MQ Series), and Broker Manager Agent with and without optimization. 

\figurename~\ref{AdaptiveSecurityManager} shows the Adaptive Security Manager. It contains the adaptive policy manager, adaptive authentication manager, adaptive authorization manager, and a KMF server. It also subscribed to information that was published from the security monitor to make changes based on the monitor’s metrics. \figurename~\ref{GEMOMTestApplication} shows a client that uses GEMOM to authenticate itself onto the network (in this case, the GEMOM authentication module used its Identity Provider and Microsoft Cardspace). Clients uses the KMF to publish or get keys for topics and clients publications or subscriptions are authorized from security tokens issued from the adaptive authorization manager. The adaptive authorization manager makes its decisions based on the current state of the network, their identity, the policy, or a combination of these.

\subsection{Validation of Results}

The GEMOM results have been validated in five real-world case studies: a collaborative business portal, a dynamic linked exchange, a financial market data delivery system, a dynamic road management system, and a banking scenario (money transfer). The results of the five case studies have enabled us to predict how the GEMOM system as a whole performs in different real-life scenarios. Below is a quick summary of the case studies.

\textbf{Collaborative business portal:} This case study was a collaborative business portal for emergency services that delivered a mechanism for cross boundary communication of critical actions in facilitating communication between teams within the emergency planning and response community. The key validation scenarios were for adaptive authentication and authorization, high dependability through broker mirroring and provision of redundant paths, and guaranteed message delivery within the workflow. A more detailed explanation is presented in \citet{Ristau2010-5628781}.

\textbf{Dynamic linked exchange:} This case study involved facilitating the economic and commercial information sharing of
\begin{enumerate*}[label=(\textit{\roman*})]
\item known procurement requirements for goods and services, principally from local government, government agencies, and SMEs;
\item availability of goods, services, and skills from SME suppliers to local government and government agency departments.
\end{enumerate*}
The validation scenarios involved delivering a “matching” service linking together procurers and suppliers with a rating mechanism to help procurers make initial choices. The validation scenarios also stress-tested the following features: system scalability, resilience, and message delivery confidentiality. A more detailed explanation is presented in \citet{Ristau2010-5628781}.

\textbf{Financial market data delivery system:} This case study involved financial market data and trading signal delivery that focused on the implementation of a real-time decision support service for institutional and private investors. The application architecture and platform analysis has been carried out with the help of selected stakeholders, i.e. institutional investors. The validation scenarios were
\begin{enumerate*}[label=(\textit{\roman*})]
\item the high-frequency data load of market information (e.g., stock quotes, foreign exchange prices or derivative quotes) to be distributed, processed and analyzed in real time,
\item the extreme variability of data load during the course of the day caused by cyclic data volumes of the tick data and the varying number of clients connected to the system, and
\item  the ultra-scalability, performance, and functional resilience.
\end{enumerate*}
A more detailed explanation is presented in \citet{Ristau2010-5628781}.

\textbf{Dynamic road management system:} This case study involved designing, developing, and testing the GEMOM self-healing and fault-tolerance on highway toll data management and collection~\citep{Paganelli2011-inproceedings}. This case study had two main scenarios:
\begin{enumerate*}[label=(\textit{\roman*})]
\item feature validation in defining reference requirements for a set of self-healing and fault-tolerance GEMOM features and validation through experimenting and testing and
\item exploitability in a specific market sector that involved a representation of a real-world application scenario.
\end{enumerate*}
This study provided valuable insights on the exploitability of GEMOM in the road transport market sector. The validation test focused on the experimentation of mirroring and self-healing capabilities of GEMOM, testing interoperability with JMS, and configuring contentment transformation rules. The test results showed that GEMOM performed well with 99.5\% correctly received messages and 5000 messages per second throughput. \Citet{Paganelli2011-inproceedings} have more detailed information.

\textbf{Universal Banking Hub:} This case study involved the placing of a Universal Banking Hub in a central architectural position as a pervasive pivoting component of the bank’s IT architecture for exchanging – both internally and externally – several types of messages using GEMOM where each message represents a specific kind of business fact~\citep{Blasi2010-10.1145/1842752.1842792}. The specific banking scenario selected utilized GEMOM to create, monitor, approve, route, track, and execute money transfers. Since high standards of security, resilience, and adaptation are required in such banking application scenarios, the focus of the test validation was on the analysis of the applicability of security metrics for adaptive authentication, authorization, end-to-end confidentiality, and the applicability of trust metrics. The validation results showed that adaptive security solutions driven by security metrics are applicable in the deployment of a Universal Banking Hub System, and increase the flexibility and security of the system by adapting to changes in the environment in accordance with the requirements of stakeholders. \Citep{Blasi2010-10.1145/1842752.1842792} have more detailed information.

\section{Discussion}

Our most important result is the development of an adaptive and evolving security system, and an adaptive trust management approach to autonomous messaging middleware systems. The model contributes to
\begin{enumerate*}[label=(\textit{\roman*})]
\item the autonomous adjustments of the run-time con-figuration of the system for preserving and maintaining optimal and uninterrupted operation,
\item the improvement of the strength of security and degree of trust in the system,
\item the improvement of the assessability and verifiability of the trustworthiness of the system, and
\item the adaptive integration of the GEMOM solution that consists of a continuous cycle of monitoring, measurement, assessment, optimization, self-healing, adaptation, and evolution to meet the challenges in the changing environments.
\end{enumerate*}  

Goals of the adaptation (self-healing, self-optimizing, self-protecting) are adapting topology, resource usage, ``fidelity,'' etc. The self-healing capabilities can prevent and recover from failure by automatically discovering, diagnosing, circumventing, and otherwise recovering from failures that might cause service disruptions. The self-optimizing capabilities enable the system to continuously tune itself both proactively to improve on existing processes and reactively in response to environmental conditions. Its self-protecting capabilities enable the system to detect, identify, and defend against viruses, unauthorized access, and denial-of-service attacks~\citep{Chess5386832}. The ensuing sections discuss this and some of the novel issues raised during the development. 

\subsection{Self-Protection through Defense In-Depth}

\Citet{IBM2005} defines a self-protecting system as a system that can anticipate, detect, identify, and protect itself against threats, unauthorized access, and denial of service attacks. GEMOM as an autonomic MOM has to implement self-protecting capabilities that can detect hostile behaviors as they occur and take corrective actions to make the system less vulnerable. The proactive identification of, and protection from, arbitrary attacks are achieved via the combination of anomaly-based self-protection and security monitoring and measurement. In our solutions, the self-protection is managed either at a single entry point (a micro property) that gives each node authorization by a coordinated defensive group attack of the other nodes alone (a macro property), or by a combination of the two (defense-in-depth). In the latter, our self-protecting system is a layered security system that manages security risks with multiple defensive strategies; that is, if one layer of defense proves inadequate, another layer of defense will prevent a full breach by containing the attack (see Section~\ref{sec:self-protection}). The different layers are as follows:

\textbf{Communication or network level defense (perimeter defense).} This consists of network level self-protection mechanisms, network level trust management scheme, mechanisms for confidentiality and integrity or authenticity of the underlying IP-network using TLS/SSL connection between routing nodes, trust models that assess the quality of new joining nodes and the degree of confidence in their behaviors, and anomaly-based Self-Protection.

\textbf{Broker nodes level defense.} This consists of trusted execution environment for nodes, and node self-protection such as mutual authentication and authorization of broker nodes for accurate namespace resolution to protect against threats from rogue brokers and to protect confidentiality and integrity.

\textbf{Publisher and subscriber level defense.} This consists of security contracts or service level agreements, use of authentication and sub-set of mechanisms to enforce access control for authorized publishers and subscribers, node-level trust management schemes (either certificate- or token-based), and adaptation and maintenance of the trust level over time by building a reputation feedback mechanism.

A key component in self-protection is the integration of mechanisms to support the detection of anomalies such as high message rates, degradation of broker performance, e.g., in the context of DoS, and of services to support the detection of anomalous message content in appropriate cases. Detectors are divided into different functions e.g., link-state detection, message-rate computation, bottleneck detection, and overall system representation~\citep{Abie2010SelfhealingAS}. The component being integrated currently uses Markov models to predict the values of different individual measurable resources of the broker (broker CPU, message rate, subscription rate, etc.) and uses a Naïve Bayes classifier, trained on system operational data, to detect a bottleneck based on the predictions~\citep{Wang2008-10.1145/1463342.1463350}. Experiments based on DoS attacks are being created and the detection and reaction mechanisms are being validated~\citep{Ristau2010-5628781}.

If human interaction is needed for interpretation, visualization of security evidence has proven to be a useful tool to increase the quality of interpretation. For example, large security metrics models are difficult to understand without visualization approaches sup-porting the simultaneous viewing of detailed measurements and higher-level objectives.

\subsection{Adaptation}

Some of the challenging issues that are discussed in the literature about adaptation include ways to cause the adaptation to occur in a running system, designing component and systems so that they can be dynamically adapted, and what to do if something goes wrong during the process of adaptation.

One issue with adaptation is the time it takes for the adaptation engine to notice an is-sue and adapt to it. Part of this delay is from the monitoring system, but part of this is also the time that the adaption engine must take to run its algorithms. New algorithms and hardware can help the engine make decisions quicker, but there will always be some sort of delay. Another way to reduce this time is to dedicate more of the overall resources to monitoring and decision-making. On the other hand, this may reduce the responsiveness and performance of the overall system. Another way to tackle the problem would be to have the adaption engine be a bit more proactive by enforcing stricter security measures before there actually is a need for them. Yet, this simply moves the window to an earlier point. The question is if the chance for an attack succeeding during this decision process time is short enough to be an acceptable risk.

Adaptation engines running on desktops and servers work fine, but it might be a problem with devices with smaller CPUs and limited battery. The Internet of Things will result in many portable devices connected to the Internet and exchanging information. On the other hand, the influx of new devices will benefit from adaptation since the system should be able to adapt to the different capabilities and requirements of all the things that are in the system instead of simply excluding them.

\subsection{Security Metrics}

Most of the security metrics efforts have been focused on the development of solutions that will be widely accepted, but lack means to obtain evidence of the security level of security-enforcing mechanisms and methodologies to relate the metrics to security objectives. Our security metrics development approaches are most valuable in the management of adaptive security and trust management, focusing on the security-enforcing mechanisms, the establishment and maintenance of trust and the quality of the overall security of the system, through sufficient and credible evidence gathering. It can be argued that some metrics require long-term data gathering before they can be utilized for adaptive monitoring and management and difficulties in attaining measurements, outlining the scope, and integrating the metrics. Our framework for assessing and calculating the trustworthiness of the development of measurable security that combines risk-based assessment of basic measurable components, a security-based trust model, and a trust-based security model into one framework extends the capabilities of each model and leverages their best features to support the adaptive development of quantifiable or measurable security. Although the overall systematic security metrics development method is an initial solution, metrics resulting from the application of the method have been utilized in the case studies that have proven to be useful in evidence based decision-support in runtime adaptive security and trust management. Further experimentation in practical situations is clearly needed to assess the feasibility of the method in the long run. The challenges include especially security risk prediction capabilities, and operational metrics adaptation in highly dynamic threat and other situations.

\subsection{Resilience and Self-adaptive Properties}

The development of our adaptive security and trust management for an autonomous messaging system – self-healing and secure self-adaptive messaging middleware is inspired by the work of many researchers~\citep{Abie2010SelfhealingAS}, but is focused more on providing resilience, self-healing, self-adaptive, integrated vulnerability management, better integration of distributed business-critical systems, and holistic and systematic adaptive security monitoring and measurement. The GEMOM system achieves a considerable increase in the end-to-end resilience of complex distributed business-critical systems to ensure secure transmission of data and services across heterogeneous infrastructures and networks~\citep{Abie2010SelfhealingAS}. The GEMOM platform consists of these resilience and self-adaptive properties:
\begin{enumerate*}[label=(\textit{\roman*})]
\item reliability of message sourcing and delivery,
\item scalability in messaging,
\item  replication of structural and dynamic properties of security policies with adaptive authentication and authorization model,
\item process-zoning and overall encapsulation to an arbitrary level, and
\item new techniques and tools for preemptive and auto-mated checking a deployed system for robustness and vulnerabilities to faults, oversights and attacks.
\end{enumerate*}
It supports a messaging infrastructure that enables adaptive functions and assurance against security vulnerabilities and erroneous input vulnerabilities to improve the reliability, robustness, and dependability of business-critical infrastructures~\citep{Wang2010TowardsAR,Abie2010SelfhealingAS}. It provides autonomous adjustments of the run-time configuration to preserve and maintain optimal and uninterrupted operation, improvement of the strength of security and degree of trust in the system, and improvement of the assessability and verifiability of the trustworthiness of the system.

\subsection{Formal Verification and Assessment}

One major challenge in adaptive systems is to provide guarantees about the required runtime quality properties. Formal methods provide the means to rigorously specify and reason about the behavior of adaptive systems. Formal methods have been applied during both system development and runtime to provide guarantees about the required properties of self-adaptive systems~\citep{Magee2006-10.1145/1137677.1137684,tamura2013towards,Weyns2012-10.1145/2168260.2168268}. The formal specification, assessment, and verification of ASM and ATM model thus involve verifying that the description of the security and trust management model ensures the correctness of security solutions. It would have been desirable to formalize and validate our proposed model using appropriate formal methods of verification and assessment, but we opted to leave it as future work.

\section{Related Work}

\subsection{MOM Platforms}

MOM platforms are available in a wide range of implementations such as JMS, Web-SphereMQ, TIBCO, Herald, Hermes, SIENA, Gryphon, JEDI, and REBECCA. Each of these MOMs has been designed to achieve specific goals, and employs unique functionality to meet specific messaging challenges~\citep{Curry2008article}. Yet, the current state-of-the-art technologies do not allow security mechanisms to actually predict or anticipate future threats nor to adapt to rapidly changing behaviors and threats over time. There are some areas of research that are promising in this regard.

\subsection{Security Measurement and Monitoring}

A number of adaptive security systems have been developed recently supporting adaptation at different levels (from hardware-level to application-level) and for a number of reasons. That is, security in an autonomic computing environment~\citep{Chess5386832}, adaptive security for wireless networks~\citep{Hager2004} and complex information systems~\citep{Shnitko2003-1222606}, adaptable security manager for real-time transactions~\citep{Son2000853993}, dynamic authentication for networked applications~\citep{schneck1998dynamic}, adaptive firewall architecture~\citep{zou2002architecture}, self-contained object for secure information distribution systems~\citep{Abie2004-10.1007/978-3-540-30191-2_42, Abie2004-article}, adaptive security policies~\citep{Lamanna2002article}, a bio-inspired self-protecting organic message-oriented middleware~\citep{Pietzowski2006-10.1007/11822035_17}, anomaly-based self-protection against network attacks~\citep{qu2006anomaly}, and virtualized trusted computing platform for adaptive security enforcement~\citep{Djordjevic2007}. Several taxonomies have been introduced for classifying adaptive and reconfigurable systems~\citep{mckinley2004taxonomy}. A survey of approaches to adaptive application security and adaptive middleware can also be found in \Citet{Elkhodary2007-4228616} and \Citet{sadjadi2003survey}, respectively. A bus-based architecture for integrating security middle-ware services is proposed in \citet{Goovaerts2008-10.1145/1463342.1463346}. Presentations of semantic and logical foundations of and local and global requirements in an adaptive security infrastructure can be found in \citet{Marcus2001, Marcus2003}. \Citet{Reinecke2010-10.1016/j.peva.2009.12.001} propose a framework and methodology for the definition of benefit-based adaptivity metrics that allow an informed choice between systems based on their adaptivity to be made, and provide a broad survey of related approaches that may be used in the study of adaptivity and to evaluate their respective merits in relation to the proposed adaptivity metric. It was the work of the above researchers that convinced us of the viability of adaptive security and trust, and therefore confidence in the productivity of our research in these directions.

\Citet{weise2008security} presents a security architecture and adaptive security, and discusses a new perspective on the characteristics of a security architecture that is capable of reducing threats and anticipating threats before they are manifested. This architecture is similar to our AES, but our AES goes further by the integration of a continuous cycle of monitoring, assessment, and evolution, and tools and processes for pre-emptive vulnerability testing and updating. The architecture is similar to ours in that it uses biological and ecosystem metaphors to provide interesting parallels for adjusting and responding to constantly emerging and changing threats, but ours goes further by combining a compromised-based trust model to maximize the value of risk-taking. 

\Citet{Wang1997TOWARDSAF} introduced a generic framework for security measurement based on a decomposition approach. \Citet{Heyman2008-4529474} utilized a security objectives decomposition approach to define a security metrics framework utilizing security pat-terns. Both approaches share similar methods with the security metrics development introduced in the present study. The Common Vulnerability Scoring System (CVSS)~\citep{Mell2006article} aims at an open and standardized method for rating vulnerabilities. The CVSS, along with some other security rating methods, has been integrated by the NIST into Security Content Automation Protocol (SCAP)~\citep{Barrett2009}. While CVSS and SCAP are important standardization efforts, their basis is not complete, and evidence of the strength of security-enforcing mechanisms is lacking. Despite several major attempts to standardize security evaluation and certification metrics, they have only achieved limited success. This is due to the fact that standards are rigid and created for certification and carrying out these processes requires a lot of time and effort. The most widely used of these efforts is the Common Criteria (CC) ISO/IEC 15408 International Standard~\citep{ISOIEC}. \Citet{Ankrum1581287} did work in security measurability with a focus on software development. \Citet{Pham2008-4561341} suggested the use of attack graphs and anomaly detection metrics. However, security effectiveness is not addressed in these contributions. Surveys of security metrics can be found from \Citet{Herrmann2007-10.5555/1204822}, \citet{Jaquith2007-10.5555/1214710}, and \citet{Bartol2009}. \Citet{Bayuk10.5555/2520890} investigated the different types of validity criteria for security metrics.

Most adaptive security monitoring approaches are still at theoretical abstraction level and are being based on Bayesian Networks or Markov chains. However, there are some practical approaches. \Citet{Ciszkowski2008-10.2478/v10065-008-0018-0} introduced an end-to-end quality and security monitoring approach for a Voice-over-Internet Protocol (VoIP) service, providing adaptive QoS and DoS/DDoS attack detection. However, this solution does not allow as much flexibility and scalability as the GEMOM Monitoring Tool. \Citet{Jean2007-4427641} described a distributed and adaptive security monitoring system based on agent collaboration. \Citet{spanoudakis2007towards} introduced a runtime security monitoring system based on confidentiality, integrity, and availability patterns. Their architecture contains a Monitoring Manager that takes requirements as an input and control Monitoring Engine. Their architecture can be directly mapped to the GEMOM monitoring architecture.

\subsection{Trust Management}

\Citet{shrobe2000active} describes an active trust management for autonomous adaptive survivable systems. The trust model described there forms the basis for our compromised-based trust model. However, ours goes further by combining this model with a security-based trust model using an adaptive control loop to minimize the rate and severity of compromises via the provision of a secure communication environment. A number of trust management systems have been developed recently that include adaptive trust negotiation and access control~\citep{Ryutov2005-10.1145/1063979.1064004}, trust-based security~\citep{Boukerche2008-10.1016/j.comcom.2008.05.007,Moloney2005-1588309}, and trustworthiness assessment~\citep{Ma2009-5336917, Luo2009-10.1016/j.ins.2009.01.039}.

\subsection{Risk Based Security}

\Citet{MCGRAW2009} describes a Risk Adaptable Access Control (RAdAC). The model is similar to our risk-based security as its access decision is based on a computation of security risk and operational need, but ours goes further by combining trust-based security and security-based trust. The effect of this combination improves and increases the strength of the security and the degree of trust in our system; it also reduces the rate and severity of compromises. The RAdAC model is similar to ours in that it considers multiple factors to determine the security risk and operational need of each access decision, but ours goes further by integrating a continuous cycle of monitoring, assessment, and evaluation and providing tools and processes for pre-emptive vulnerability testing and updating. This integration improves and increases the assessability and verifiability of the trustworthiness of the system. A number of Risk-based security and trust management have recently been developed that include threat-adaptive security policy~\citep{Venkatesan581559,Savola2010article}.

\section{Conclusion and Future Perspectives}

In this paper we have described GEMOM, which provides solutions to overcome limitations in robustness, resilience, adaptability and scalability, and have presented an adaptive and evolving security (AES) and an adaptive trust management (ATM) approach to such autonomous MOM systems, an approach that is capable of (a) maintaining the proper balance between security and trust, and performance in rapidly changing environments, (b) improving and increasing the strength of security and degree of trust in the system by combining adaptive risk-based security, trust-based security, and security-based trust, and (c) improving the ability to assess and verify the trustworthiness of the system by integrating different metrics, assessment and observation tools. We have also analyzed the theoretical foundations of adaptivity, upon which our models are based, with its benefits and shortcomings, and introduced a framework for the assessment of trustworthiness and calculation of confidence. 

Our investigations convince us that our AES and ATM models of GEMOM are capable of impacting on the robustness of communication between users with disparate devices and networks, secure self-healing and self-adaptive systems that support mission-critical communication under highly dynamic environmental conditions, self-auditing systems that can report the state inconsistencies, and incorrect or improper use of components, systematic secure evolution of legacy software to accommodate new technologies and adapt to new environments, and of enabling systems to operate in the face of failures and attacks.
Utilization of risk-driven security metrics in adaptive security management is a promising approach. It should be noted that the development of metrics requires rigorous development of security objectives and suitable mechanisms to support sufficient availability and attainability of relevant security effectiveness evidence.

The results of the five case studies have enabled us to predict how the GEMOM system as a whole performs in different real-life scenarios. The validation results showed that adaptive security and adaptive trust management solutions driven by holistic systematic security monitoring and measurement, and security metrics are applicable in the deployment of distributed business-critical systems, and increase the flexibility and security of the system by adapting to changes in the environment in accordance with the requirements of stakeholders.

In our future work we plan to improve the self-abilities of AES and ATM components to detect in real time unknown privacy threats, to swiftly respond to them, and to adapt to the dynamism of the environment and to the changing degree of privacy breaches. Finally, we intend to enhance our AES and ASM framework with a formal model to rigorously verify and reason the correctness of security and trust management solutions and the behavior of self-adaptive system.

\begin{acks}
This research is in the context of the EU project GEMOM (Genetic Message-Oriented Secure Middleware), grant agreement no 215327, approved by the EU Commission. The authors acknowledge the efforts of all partners who participated in the development of the GEMOM framework.
\end{acks}

\bibliographystyle{ACM-Reference-Format}
\bibliography{main}

\end{document}